\documentclass[traditabstract]{aa} 
\usepackage{graphicx}
\usepackage{txfonts}
\begin{document}
   \title{Can angular momentum loss cause the period change of NN Ser?}

   \author{Wen-Cong Chen          }

   \institute{Department of Physics and Information Engineering, Shangqiu Normal University, Shangqiu 476000, PR China\\
              \email{chenwc@nju.edu.cn} }

   \date{}


  \abstract
{NN Ser is a non mass-transferring pre-cataclysmic variable
containing a white dwarf with a mass of $\sim 0.5 M_{\odot}$ and
an M dwarf secondary star with a mass of $\sim 0.2 M_{\odot}$.
Based on the data detected by the high-speed CCD camera ULTRACAM,
it was observed that the orbital period of NN Ser is decreasing,
which may be caused by a genuine angular momentum loss or the
presence of a third body. However, neither gravitational radiation
and magnetic braking can ideally account for the period change of
NN Ser. In this Letter, we attempt to examine a feasible mechanism
which can drain the angular momentum from NN Ser. We propose that
a fossil circumbinary disk (CB disk) around the binary may have
been established at the end of the common envelope phase, and the
tidal torques caused by the gravitational interaction between the
disk and the binary can efficiently extract the orbital angular
momentum from the system. We find that only if M dwarf has an
ultra-high wind loss rates of $\sim 10^{-10} M_{\odot} \rm
yr^{-1}$, and a large fraction ($\delta\sim 10 \%$) of wind loss
is fed into the CB disk, the loss rates of angular momentum via
the CB disk can interpret the period change observed in NN Ser.
Such a wind loss rate and $\delta$-value seem to be incredible.
Hence it seems that the presence of a third body in a long orbit
around the binary might account for the changing period of NN Ser.
   }

   \maketitle


\section{Introduction}

From the Palomar-Green survey, NN Ser (PG 1550+131) was observed
to have the clear emission lines in its spectra. Based on the
broad H, He I profiles, and Balmer jumps in the emission lines, NN
Ser was classified as U Gem-like object (\cite{gree82}). Through
the observations for NN Ser, \cite{wils86} found that there exists
a rapid decline in the brightness of $\sim 0.6$ magnitudes, which
was believed to be a characteristic of cataclysmic variable stars.

Based on the photometric observations during the nights of 1988
July, there exhibits a strong heating effect ($\sim 0.6$ mag) and
very deep primary eclipses ($\ga0.4$ mag), which show that NN Ser
is a detached white dwarf binary with an orbital period of 3.12
hr, namely it is a pre-cataclysmic variable (\cite{haef89}).
Through the low-resolution \emph{IUE} observations for NN Ser,
\cite{wood91} derived that the mass of the white dwarf is in the
range of $0.47 - 0.60 M_{\odot}$, and the secondary star is an M
4.7-M 6.1 dwarf with a mass of $0.09 - 0.14 M_{\odot}$ and a
temperature of $2775 - 3050$ K. Using the phase-resolved blue and
far red spectra, \cite{cata94} have inferred the radial
velocities, and discussed the physical and orbital parameters for
both components of NN Ser. Recently, \cite{haef04} redetermined
all system parameters of NN Ser by using the VLT and the
multi-mode FORS instruments.

Using the high speed CCD camera ULTRACAM, \cite{brin06} have
detected 13 primary eclipses of NN Ser. After fitting of light
curve models, they concluded that the orbital period of the binary
is decreasing, with a mean change rate $\dot{P}=(9.06\pm
0.06)\times 10^{-12}\rm ss^{-1}$, and a current change rate
$\dot{P}=(2.85\pm 0.15)\times 10^{-11}\rm ss^{-1}$ during the past
2 years. Generally, period changes in binary systems may be caused
by three mechanisms: Applegate's mechanism (\cite{appl92}), the
presence of a third body around the binary, and an orbital angular
momentum loss. After excluding Aplegate's mechanism, \cite{brin06}
concluded that a third body with a mass of $0.0043 - 0.18
M_{\odot}$ and an orbital period of $30 - 285 $ yr may be
responsible for the period change of NN Ser. Furthermore, the
standard magnetic braking model proposed by \cite{rapp83} can also
account for NN Ser's period change if the magnetic braking
mechanism still works at a secondary star with mass lower than
$0.3 M_{\odot}$ (\cite{brin06}).

Grossman, Hays \& Graboske (1974) suggested that main-sequence
stars with masses $\la 0.3 M_{\odot}$ have convective cores,
therefore the magnetic field lines can not be locked to the core
and magnetic braking should be cut off. From the observed angular
momentum loss properties of cataclysmic variables (CVs),
\cite{andr03} noted that there is no evidence for a cut-off in
magnetic braking when the stellar mass is below $0.3 M_{\odot}$.
In addition, if gravitational radiation is the unique mechanism
draining angular momentum from CVs with a short period, it would
result in two serious consequences: (1) the inferred minimum
orbital period would be at 1.1 hr instead of the observed value of
1.3 hr; (2) there should exist a significant fraction of CVs with
a minimum period, which is not consistent with the observations
(\cite{patt98}).  Therefore, it is a controversial issue whether
magnetic braking still works for NN Ser.

It is the purpose of this paper to examine a feasible mechanism
extracting the orbital angular momentum from NN Ser, and present
an alternative interpretation for its orbital period change. In
subsection 2.1, our analytic result indicate that the magnetic
braking can not interpret the observed data of NN Ser. In section
2.2 we investigate if a circumbinary (CB) disk around the binary
system can be responsible for the period change of NN Ser.
Finally, we make a brief discussion and summary in section 3.

\section{The orbital angular momentum loss}
\subsection{Magnetic braking}
Single low-mass main-sequence stars would be expected to undergo
magnetic braking due to the coupling between the stellar winds and
the magnetic field (\cite{verb81}). The specific angular momentum
loss in the stellar winds is very large because the outflow
material is tied in the magnetic field lines to co-rotate with the
stars out to a long distance (\cite{webe67,mest87,kalo99}).

Assuming that magnetic braking still works for NN Ser, we make an
estimation for the loss rates of orbital angular momentum. The
magnetospheric radius of M dwarf can be derived as (\cite{just06})
\begin{equation}
r_{\rm m}=(GM_{\rm d})^{-1/8}B_{\rm s}^{1/2}R_{\rm
d}^{13/8}\dot{M}_{\rm wind}^{-1/4},
\end{equation}
where $G$ is the gravitational constant, $B_{\rm s}$ is the
surface magnetic field of the star, $M_{\rm d}$ and $R_{\rm d}$
are the mass and the radius of the star, respectively,
$\dot{M}_{\rm wind}$ is the stellar wind loss rates.

In a binary system, the angular momentum loss by magnetic braking
would cause the evolved star to spin down. However, the tidal
forces between the two components would continuously act to spin
the star up back into co-rotation with the orbital rotation
(\cite{patt84}). The spin-up takes place at the expense of the
orbital angular momentum. Hence magnetic braking indirectly
carries away the orbital angular momentum of the binary system.
Assuming that the stellar wind with the angular velocity of the
donor star at magnetosphere depart from the magnetic lines, the
loss rates of orbital angular momentum is given by Justham,
Rappaport \& Podsiadlowski (2006)
\begin{equation}
\dot{J}_{\rm mb}=-2\pi P_{\rm orb}^{-1}(GM_{\rm d})^{-1/4}B_{\rm
s}R_{\rm d}^{13/4}\dot{M}_{\rm wind}^{1/2},
\end{equation}
here $P_{\rm orb}$ is the orbital period of the binary .

For NN Ser, the mass and radius of the M dwarf are
$\sim0.15M_{\odot}$ and $\sim0.18R_{\odot}$, respectively
(\cite{haef04}). Detection of Zeeman splitting of Fe line in the
very active M dwarfs indicates a magnetic field of 2-4 kG
(\cite{john96}). Based on the analyzing data from the James Clerk
Maxwell Telescope, \cite{mull92} proposed that the flaring M dwarf
stars have a ultra-high wind loss rates of $\sim
10^{-10}M_{\odot}\rm yr^{-1}$. Hence we can obtain the loss rates
of orbital angular momentum by magnetic braking
\[
\dot{J}_{\rm mb}=-5.6\times 10^{34} \left(\frac{P_{\rm
orb}}{0.13\rm d}\right)^{-1}\left(\frac{M_{\rm
d}}{0.15M_{\odot}}\right)^{-1/4}\left(\frac{B_{\rm s}}{4000\rm
G}\right)
\]
\begin{equation} \left(\frac{R_{\rm
d}}{0.18R_{\odot}}\right)^{13/4}\left(\frac{\dot{M}_{\rm
wind}}{10^{-10}M_{\odot}\rm yr^{-1}}\right)^{1/2}\rm
g\,cm^{2}\,s^{-2}.
\end{equation}
Since $10^{-10}M_{\odot}\rm yr^{-1}$ is indeed the upper limit of
the wind loss rate for a M dwarf (see subsection 2.2), $5.6\times
10^{34}\rm g\,cm^{2}\,s^{-2}$ is an absolute maximum of $\dot{J}$.
Apparently, Eq.~(3) shows that magnetic braking is not likely to
account for the period change of NN Ser, in which the average
angular momentum loss rates is in the range of $0.84-2.09\times
10^{35} \rm g\,cm^{2}\,s^{-2}$ (\cite{brin06}). The standard
magnetic braking model proposed by \cite{rapp83} overestimates
angular momentum loss rates for low-mass stars with high rotation
rates. However, with the saturated magnetic braking by
\cite{sill00} and \cite{andr03}, the rate of angular momentum loss
is too low to explain the period change in NN Ser (\cite{brin06}).
Hence there should be a more efficient angular momentum loss
mechanism to explain the current orbital evolution of NN Ser.

\subsection{Circumbinary disk}
Recently, a new mechanism to extract angular momentum from binary
stars --- CB disk --- was noticed by many authors. The idea about
CB disks was firstly proposed by \cite{heuv73} and \cite{heuv94},
and they argued that part of the transferred material with a large
orbital angular momentum may form a disk around the binary rather
than leaving it. The rotating disk around a binary is called CB
disk, whose schematic diagram was presented by \cite{haya09} (see
their Figure 1). If the CB disk follows Keplerian rotation, its
angular velocity $\Omega \propto r^{-3/2}$ ($r$ is the distance
between the CB disk and the mass center of the binary system),
then the angular velocity of the CB disk is less than that of the
binary. The gravitational interaction between the inner edge
$r_{\rm i}$ of the CB disk and the binary causes the disk to spin
up, and causes the binary to spin down. Hence the tidal torques
caused by the gravitational interaction may extract the orbital
angular momentum from the binary to the CB disk. \cite{spru01} and
\cite{taam01} invoked the CB disk to explain a large spread of the
mass transfer rate for a given orbital period in CVs. Our recent
work also shows that CB disks can efficiently extract orbital
angular momentum from binaries, and hence enhance the mass
transfer rates (\cite{chen06a,chen06b,chen07}).

In the standard scenario for the birth of cataclysmic variable,
the progenitor systems were assumed to include an intermediate
mass star and a low mass companion. After the more massive star
fills its Roche lobe, the mass transfer rate is very high because
of the large mass ratio between the two components, and the binary
systems will evolve into a common envelope phase (for a review
\cite{iben93}). With ejecting the envelope due to friction
dissipation, a compact binary with a low mass white dwarf was
formed. NN Ser was proposed to be a post common envelope binary
(\cite{schr03}). Since the common envelope may not be entirely
ejected, a fossil CB disk may have been established
(\cite{spru01}).

Assuming that a fraction $\delta$ of the stellar wind loss from
the M dwarf feeds into the CB disk, the loss rates of angular
momentum via the CB disk can be written as (\cite{spru01,taam01})
\begin{equation}
\dot{J}_{\rm cb}=-\gamma\left(\frac{2\pi a^2}{P_{\rm
orb}}\right)\delta\dot{M}_{\rm wind}\left(\frac{t}{t_{\rm
vi}}\right)^{1/2},
\end{equation}
where $\gamma=\sqrt{r_{\rm i}/a}$ ($a$ is the binary separation),
$t$ is the mass input timescale of the disk. Under the assumption
of the standard thin disk, the viscous timescale $t_{\rm vi}$ at
the inner edge of the CB disk is given by $ t_{\rm
vi}=2\gamma^{3}P_{\rm orb}/(3\pi\alpha_{\rm SS}\beta^{2}), $ where
$\beta=H_{\rm i}/r_{\rm i}$, $\alpha_{\rm SS}$ and $H_{\rm i}$ are
the viscosity parameter and the scale height of the CB disk,
respectively (\cite{chen07}).

For NN Ser, \cite{haef04} suggested that the cooling time of the
white dwarf is $\sim1.3\times 10^{6}$ yr, which is consistent with
the age of the white dwarf estimated by \cite{wood91}. Setting
$\gamma=1.3, \alpha_{\rm SS}=0.01$, and $\beta=0.03$
(\cite{chen06b}), $ t_{\rm vi}\approx 18$ yr. Inserting typical
values for the parameters into equation (4), we have

\[
\dot{J}_{\rm cb}=-5.5\times 10^{35} \left(\frac{P_{\rm
orb}}{0.13\rm
d}\right)^{-1}\left(\frac{a}{0.95R_{\odot}}\right)^{2}\left(\frac{\delta}{0.1}\right)
\]
\begin{equation} \left(\frac{\dot{M}_{\rm
wind}}{10^{-10}M_{\odot}\rm
yr^{-1}}\right)\left(\frac{t}{1.3\times 10^{6}\rm yr}\frac{18\rm
yr}{t_{\rm vi}}\right)^{1/2}\rm g\,cm^{2}\,s^{-2}.
\end{equation}
According to equation (5), the CB disk can explain the loss rates
of angular momentum seen in NN Ser for a large $\delta=0.1$ and an
ultra-high wind loss rate ($10^{-10}M_{\odot}\rm yr^{-1}$).

The above order-of-magnitude estimate obviously contains
substantial uncertainties in $\delta$ and the wind loss rates.
Firstly, the loss rates of angular momentum relies strongly on the
stellar wind loss rates, which may be overestimated by
\cite{mull92}. Based on the observations for several M dwarf flare
stars, an upper limit of $\sim 10^{-12}M_{\odot}\rm yr^{-1}$ was
derived (\cite{lim96,oord97}). Analysis of the data from
\emph{Chandra} and \emph{Hubble} observations showed that, the M
5.5 dwarf Proxima Centauri has a wind loss rate of $\sim
10^{-14}-10^{-15}M_{\odot}\rm yr^{-1}$ (\cite{wood01,warg02}).
Assuming that the white dwarf accretes wind material through
Bondi-Hoyle accretion, \cite{debe06} presented a wind loss rates
range of $\sim 10^{-14}-10^{-16}M_{\odot}\rm yr^{-1}$ for three M
dwarfs. We expect that future observations on P Cygni profiles,
optical and molecular emission lines, infrared and radio excesses,
and absorption lines can present further constrain for the wind
loss rates of NN Ser (\cite{lame99}). Secondly, even if the wind
loss rates derived by \cite{mull92} are the same with NN Ser, the
CB disk mechanism still requires a large $\delta$, which is about
2-3 orders of magnitude larger than that used by \cite{spru01}. In
a word, such wind loss rates and $\delta$-values are incredibly
large, so it seems that the presence of a CB disk is not the main
cause for the period change of NN Ser.

\section{Discussion and summary}
Recently, Brinkworth et al. (2006) suggested that the standard
magnetic braking model may explain the period change observed in
NN Ser if the magnetic braking cut-off was ignored. In this
letter, we estimate the loss rates of angular momentum via
magnetic braking by an analytic approach. Our result shows that,
even if M dwarf in NN Ser possesses a strong magnetic field of
$4000$ G and an ultra-high wind loss rates of $\sim
10^{-10}M_{\odot}\rm yr^{-1}$, the loss rates of angular momentum
via magnetic braking are an order of magnitude less than that of
observation. However, for the same wind loss rates, if a large
fraction ($\sim 10\%$) of wind loss was input the CB disk, the
loss rates of angular momentum seen in NN Ser can be interpreted
by the tidal torques caused by the gravitational interaction
between the CB disk and the binary.

 However, the stellar wind loss rates of M dwarfs determine
 if a CB disk can account for the period change of NN Ser.
 Several authors have subsequently derived the wind loss rates of
2-6 magnitudes lower than the one given by Mullan et al. (1992)
(\cite{lim96,oord97,wood01,warg02,debe06}). Based on recent
inferred wind loss rates for M dwarfs, the CB disk is not the main
mechanism causing the period change of NN Ser. If we can not
explore a more efficient mechanism extracting angular momentum
from the binary, the presence of a third body in a long orbit
around NN Ser may be the best candidate mechanism to cause its
period change (\cite{brin06}).

Though a CB disk is ruled out in NN Ser, the existence of CB disks
might be a key issue in studying the evolution of the CVs. As a
result of a continuum contribution of the dust emission, the CB
disk may be detected in the L waveband (\cite{spru01}). Recently,
Hayasaki \& Okazaki (2009) suggested a new channel to probe a CB
disk, in which the emission profiles may be caused to be variable
by prograde and nonaxisymmetric waves. Through interferometric
observations, the direct imaging of CB disks has been successfully
obtained in some young binaries such as GG Tau (\cite{dutr94}) and
UY Aur (\cite{duve98}). When the Spitzer data for CVs with a
strong magnetic field were analyzed, the flux density of four and
five polars in mid-infrared was discovered to be in excess,
respectively. Howell et al. (2006) and Brinkworth et al. (2007)
proposed that the CB dust disks may be the best origin of these
excess (but the source of optically thin cyclotron cannot be ruled
out). Recently, Dubus et al. (2007) found that the infrared
emission from the magnetic CVs AE Aqr is obviously larger than the
expected value from the companion, and they thought that the
thermal emission from the CB material might be a candidate.
Therefore, as the progenitor of CVs, NN Ser may also be surrounded
by a CB disk. We expect further detailed multi-waveband
observations for this pre-CVs to confirm or negate our idea in the
future.

\begin{acknowledgements}
We thank the anonymous referee for his/her helpful comments
improving  this manuscript, and thank Stephen Justham for his help
in improving English of this paper.
 This work has been supported in part by the National Natural Science
 Foundation of China (Grant No 10873011), and sponsored by Program
for Science \& Technology Innovation Talents in Universities of
Henan Province, China.
\end{acknowledgements}

\end{document}